\documentclass[graybox]{svmult}

\usepackage{mathptmx}       
\usepackage{helvet}         
\usepackage{courier}        
\usepackage{type1cm}        
\usepackage{makeidx}         
\usepackage{graphicx}        
\usepackage{bm}
\usepackage{multicol}        
\usepackage[bottom]{footmisc}

\makeindex             

\begin{document}

\title*{Maximizing a psychological uplift in love dynamics}
\author{Malay Banerjee, Anirban Chakraborti and Jun-ichi Inoue}

\institute{
Malay Banerjee \at Department of Mathematics and Statistics,
Indian Institute of Technology,
Kanpur-208016, INDIA
\email{malayscc@gmail.com} \and
Anirban Chakraborti \at School of Computational and Integrative Sciences, Jawaharlal Nehru University, New Delhi-110067, INDIA
\email{anirban@mail.jnu.ac.in}
\and Jun-ichi Inoue \at Graduate School of Information Science and Technology,
Hokkaido University, N14-W9, Kita-ku, Sapporo 060-0814, JAPAN \email{jinoue@cb4.so-net.ne.jp}
}
\maketitle

\abstract{
In this paper, we investigate the dynamical properties of
a psychological uplift in lovers.
We first evaluate extensively the dynamical equations which were recently given by
Rinaldi \textit{et. al.} (2013) \cite{Rinaldi}.
Then, the dependences of the equations on several parameters are numerically examined.
From the view point of lasting partnership for lovers, especially, for married couples,
one should optimize the parameters appearing in the dynamical equations
to maintain the love for their respective partners.
To achieve this optimization, we propose a new idea where the parameters are stochastic variables
and the parameters in the next time step are given as expectations over a
Boltzmann-Gibbs distribution at a finite temperature.
This idea is very general and might be applicable to other models dealing with human relationships.
}

\section{Introduction}

\begin{quote}
``Love is composed of a single soul inhabiting two bodies.'' -- Aristotle

``Love never dies a natural death. It dies because we don't know how to replenish its source. It dies of blindness and errors and betrayals. It dies of illness and wounds; it dies of weariness, of witherings, of tarnishings.'' -- Ana\"{i}s Nin
\end{quote}

Love -- mysterious and unexplained -- often forms the basis of a relationship between two persons; undoubtedly,
a partnership between lovers is a time-dependent phenomenon.
 Even if  a man and a woman were
 in deep love at some initial stages, the psychological uplift for one or both of them
 could eventually decay to
 very low-levels, and could even result in a break-up or divorce, in the worst scenario.

A simple mathematical model for the dynamics of love between a man and woman was introduced by
Strogatz \cite{Strogatz1, Strogatz2} -- the first attempt to model the love dynamics with the help of coupled ordinary differential equations. The idea of Strogatz was then extended by other researchers \cite{RinaldiIJBC, RinaldiAMC, RinaldiNDPLS} to understand the influence of the factors like appeal, secure relation between the couple, separation for a finite time period, which are important factors to maintain the relationship. Similar type of mathematical models have been proposed and analyzed up to certain extent for triangular love by Sprott \cite{Sprott1, Sprott2} but the uncertainty for the final outcome remains unclear.
Recently, Rinaldi \textit{et. al.} \cite{Rinaldi} again proposed a simple
dynamical model for lovers emotion to investigate a law of big hit
film from the dynamical behavior of feeling in the partner for
lovers. Their approach, based on a coupled differential
equations, was applied to the movie `Gone With The Wind' (GWTW); they found that the resulting
time series of lovers' feelings can mimic the story of the film to
some extent. The differential equations contain several parameters
and Rinaldi \textit{et. al.} chose them to mimic the lives of Scarlet and Rhett, with
full of ups and downs. 
In the romantic film GWTW,
 the drastic ups and downs in the lovers' emotions indeed constituted a notable factor to
 attract the attention of audience and the sequences of such psychological
 climaxes in the film might have been a key issue in making the film a big hit, as suggested by Rinaldi \textit{et. al.} (2013) \cite{Rinaldi}.

In reality, for a married couple,
such extreme ups and downs could however prove to be deterrent to the continuation a peaceful
married life. Hence, from the view point of lasting partnership for
lovers, especially for a married couple, one should optimize the
parameters appearing in the dynamical equations to maintain the
love for their partner. In other words, it would be interesting to obtain
the optimum levels of the parameters in order to maintain the
minimum level of love and happiness required to maintain a happy and prolonged marital life.

To this aim, we propose a simple new idea in this paper. We
assume that the parameters involved with the love dynamics are not
constant over the entire time period, rather they are stochastic
variables and the parameters in the next time step are given as
expectations over a Boltzmann-Gibbs distribution at a finite temperature. 
By decreasing the temperature during the dynamics 
of coupled equations, one can accelerate 
the rate of increase of the sum of feelings (and decrease the difference of feelings) 
of lovers at each time step. 
The idea is quite general and might be applicable to other models
dealing with human relationships.

\section{Differential equations of gross and gap for lovers' feelings}
\label{sec:1}
In the original model by Rinaldi \textit{et. al.} \cite{Rinaldi}, the governing equations with respect to the feelings of lovers, denoted as $x_{1},x_{2}$, are given by two
coupled non-linear ordinary differential equations:
\begin{eqnarray}
\frac{dx_{1}}{dt} & = &
-\alpha_{1} x_{1} + \rho_{1}A_{2} +k_{1}x_{2}\,{\rm e}^{-\eta_{1} x_{2}}, \\
\frac{dx_{2}}{dt} & = & -\alpha_{2} x_{2} + \rho_{2}A_{1}
+k_{2}x_{1}\,{\rm e}^{-\eta_{2} x_{1}},
\end{eqnarray}
subjected to the positive initial conditions, where the parameter $\alpha_i$ is {\it
forgetting coefficient}, $k_i$ and $\eta_i$ are the parameters
characterizing the measure of insecurity feelings, $A_j$ is the
measure of appeal towards $x_i$ produced by $x_j$ and $\rho_i$ is a
multiplicative factor representing the amount of recognition of the
appeal $A_j$ (see \cite{Rinaldi} for detailed interpretation). All
the parameters involved with the model are positive. Interestingly, once we choose the initial values of $x_{1},x_{2}$, these variables remain positive.

As one can see above, that there are many parameters to be calibrated.
From the engineering point of view, one could determine them
by means of `optimization' of some appropriate cost functions.
In the following, we consider several such cost functions.

First, we
introduce the following new variables, namely, the `gross' $S$ (sum) and `gap' 
$D$ (difference):
\begin{equation}
S \equiv x_{1}+x_{2},\,\,\, D \equiv (x_{1}-x_{2})^{2} =
x_{1}^{2}+x_{2}^{2}-2x_{1}x_{2}.
\end{equation}
This allows us to write:
\begin{equation}
x_{1}=\frac{1}{2}(S+\sqrt{D}), \,\,
x_{2}=\frac{1}{2}(S-\sqrt{D} ),
\end{equation}
where we should bear in mind that we have to consider the case
$x_{1} \geq x_{2}$ in order to have the well-defined expressions for $x_1$ and $x_2$ 
in terms of $S$ and $D$. Of course, this condition may not always be satisfied. However, as we are focusing here on the gap $D$, the above choice
might be indeed justified. It should be noted that the gross feelings $S$
could be regarded as a cost function to be maximized. This is
because the total degree of `passion' amongst the lovers might be one of the most
important quantities to make the relationship strong and durable. On the
other hand, the gap the two partners' love
$x_{1},x_{2}$ might determine the `stability' of the relationship -- namely, even if the $S$ is high, the mutual relation could be unstable when $x_{1} \gg x_{2}$ or $x_{1} \ll
x_{2}$. In other words, it is very hard for the lovers to continue
their good relationship if only one of them expresses too much love to
his/her love partner and the other partner becomes indifferent about their
relationship which was established due to their love affairs. Two hypothetical cases can be considered for illustrating this.
\begin{itemize}
\item
For young lovers,
the variable $S$ takes high values temporally;
however, one person (girl or boy) suddenly loses interest and becomes indifferent.
As a result,  the variable $D$ increases rapidly and the love affair (marriage) breaks down prematurely.
\item
For senior lovers,
the variable $S$ normally does not take a high value;
however, they know each other quite well, and
as a result, the feelings $x_{1}$ and $x_{2}$ are quite similar.
Hence, variable $D$ increases and the love affair (marriage) becomes stable.
\end{itemize}
We do not have any real survey data to validate these idealized examples. Nevertheless,
we consider an utility function $S$, which is to be maximized,
and the energy function $D$, which is to be minimized,
in order to determine the
parameters appearing in the original model \cite{Rinaldi}.

Then, the original equations are rewritten
in terms of $S$ and $D$.
The equation for $S$ is easy to obtain, and we have
\begin{eqnarray}
\frac{dS}{dt} & = &
-\alpha_{1} \frac{S+\sqrt{D}}{2}
+\rho_{1}A_{2} +
k_{1}\frac{S-\sqrt{D}}{2}{\rm e}^{-\eta_{1}(S-\sqrt{D})/2} \nonumber  -
\alpha_{2} \frac{S-\sqrt{D}}{2} \nonumber \\
\mbox{} & + & \rho_{2}A_{1} +
k_{2}\frac{S+\sqrt{D}}{2}{\rm e}^{-\eta_{2} (S+\sqrt{D})/2} \equiv  f(\bm{\theta}:S,D)
\label{eq:f} \\
\frac{dD}{dt}  & = &
2(x_1-x_2)\left(\frac{dx_1}{dt}-\frac{dx_2}{dt}\right)\ \equiv
g(\bm{\theta}:S,D),
\label{eq:g}
\end{eqnarray}
where $\bm{\theta} \equiv
(\alpha_{1},\alpha_{2},\rho_{1},\rho_{2},A_{2},A_{2},k_{1},k_{2},\eta_{1},\eta_{2})$.

In the following parts, we discuss in details, the behavior of the non-linear dynamics of equations
(\ref{eq:f})-(\ref{eq:g}), within the framework of Rinaldi {\it et. al.}  \cite{Rinaldi} model, and 
consider the possible optimization of the parameters. Here, we have chosen the model by  Rinaldi {\it et. al.} just as a basic example, and in principle one could easily extend the study by taking into account much more complicated and appropriate lovers' interactions. 
\subsection{Some specific choices of parameters}
\begin{figure}[ht]
\begin{center}
\includegraphics[width=8.5cm]{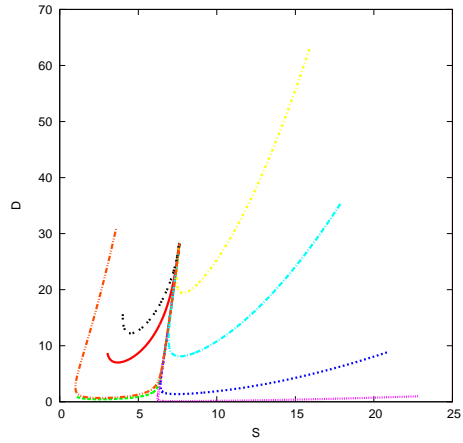}
\end{center}
\caption{\footnotesize
The flows $S$-$D$ calculated by Eqs. (\ref{eq:f})-(\ref{eq:g}) for
several distinct initial conditions. We find that for any initial conditions, the flows converge to $(7.61, 28.19)$.
We set the parameters as
$\alpha_1 =\alpha_2 =\rho_1 =\rho_2 =
\eta_1 = \eta_2 =1$ and
$k_1 =k_2 =15,
A_1 =A_2= 1$.
 }
 \label{fig:fg1}
\end{figure}

We first examine the behavior of the differential equations (\ref{eq:f}) and (\ref{eq:g}) 
with respect to $S$ and $D$ for the case of a specific choice of parameters $\bm{\theta}$.
\begin{figure}[ht]
\begin{center}
\includegraphics[width=5.8cm]{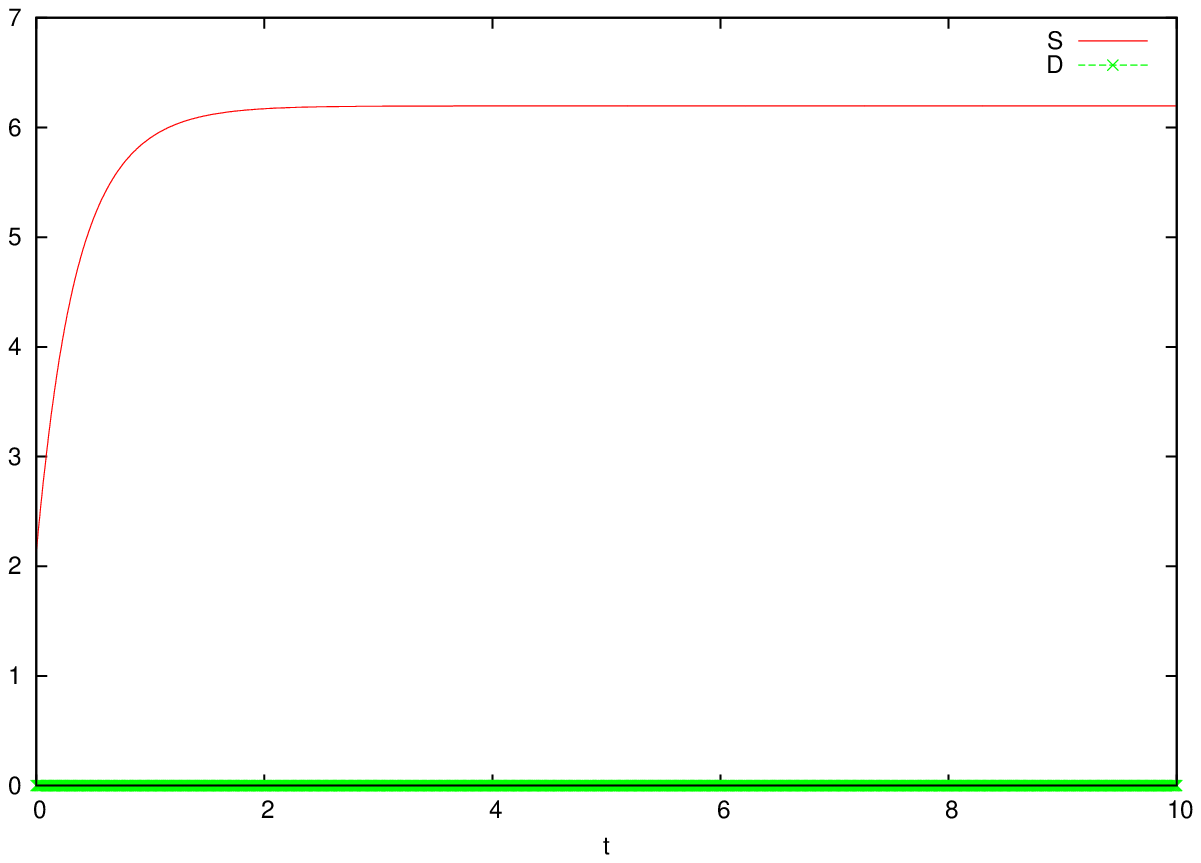}
\includegraphics[width=5.8cm]{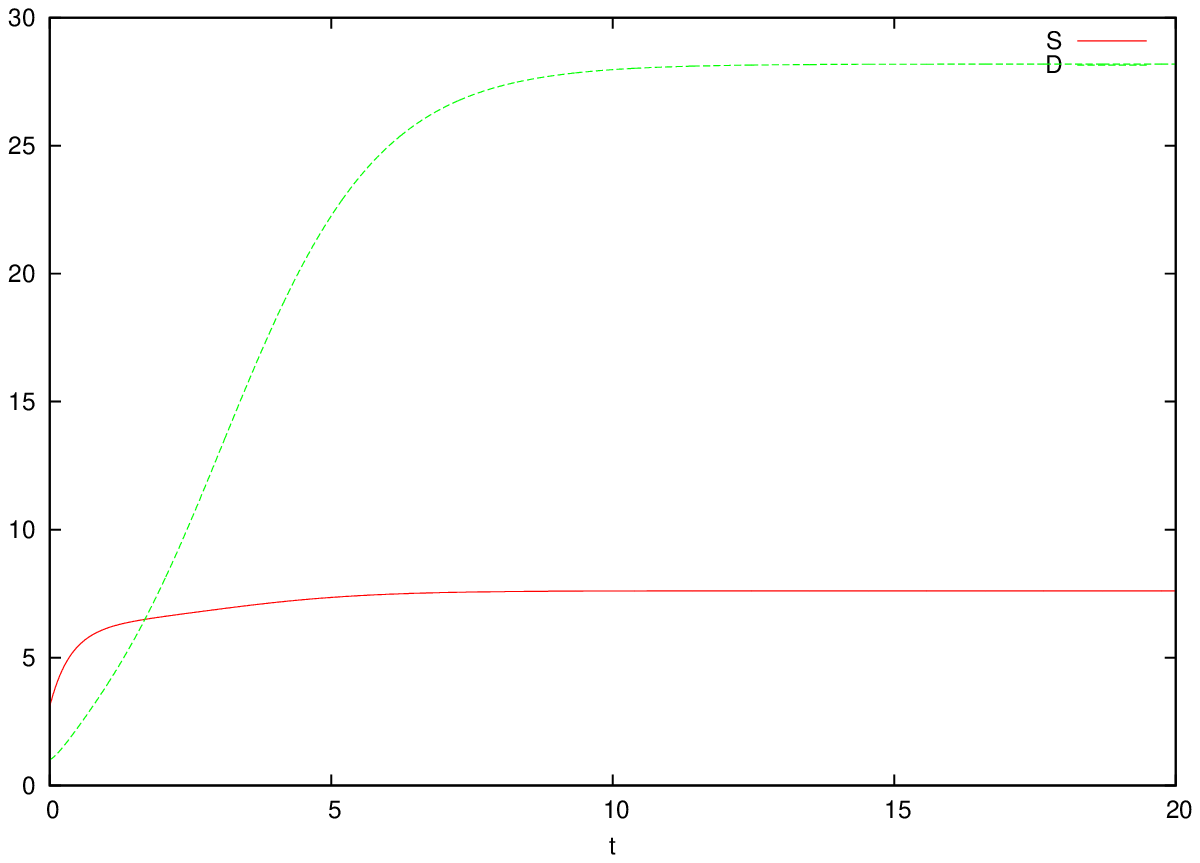}
\end{center}
\caption{\footnotesize
The time-dependence of $S$ and $D$ calculated by Eqs. (\ref{eq:f})-(\ref{eq:g}). 
We set the parameters as
$\alpha_1 =\alpha_2 =\rho_1 =\rho_2 =
\eta_1 = \eta_2 =1$ and
$k_1 =k_2 =15,
A_1 =A_2= 1$.
The initial condition are selected as
$(x_{1},x_{2})=(1,1)$ (left),
this reads $D_{0}=0$,  and
$(x_{1},x_{2})=(2,1)$ (right),
this reads $D_{0} \neq 0$.
 }
 \label{fig:fg0}
\end{figure}
Apparently, $D_{t}=0$
is always a solution of the equation (\ref{eq:g}).
In Fig. \ref{fig:fg0}, we plot
the $S_{t}$ and $D_{t}$ for
two distinct initial conditions.
In the left  panel,
we choose
the initial condition so that 
$x_{1}(0)=x_{2}(0)=1$,
this reads $D_{0}=0$.
From this panel, we easily find that
the gap $D$ is time-independently zero.
On the other hand, in the right panel, we choose
as $x_{1}(0) \neq x_{2}(0)$,
namely,
$D_{0} \neq 0$. For this case,
the gap $D$ evolves in time and
converges to some finite value.
In Fig. \ref{fig:fg1},
we show the flows (trajectories) $S$-$D$ for
$D_{0} \neq 0$.
All flows converge to $(7.61, 28.19)$.
\subsubsection{Symmetric case}
For symmetric case $D_{t}=0$ ($x_{1}=x_{2}$),
the differential equation
with respect to $S$ is simply obtained by
\begin{equation}
\frac{dS}{dt} =
-\alpha S  + 2\rho A + k S {\rm e}^{-\eta S/2}.
\end{equation}
The steady state is given by the following non-linear equation.
\begin{equation}
\alpha S = 2\rho A + kS{\rm e}^{-\eta S/2}
\label{eq:steady}
\end{equation}
\begin{figure}[ht]
\begin{center}
\includegraphics[width=5.8cm]{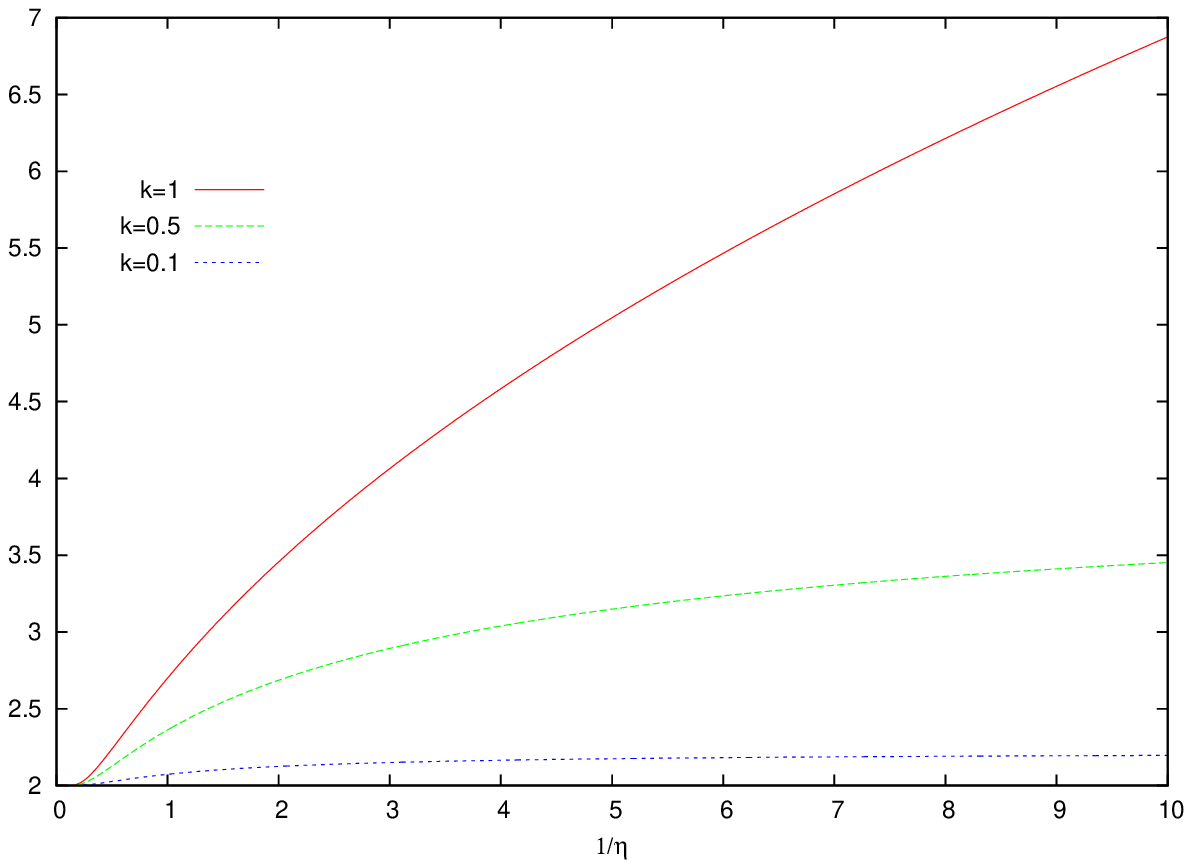}
\includegraphics[width=5.8cm]{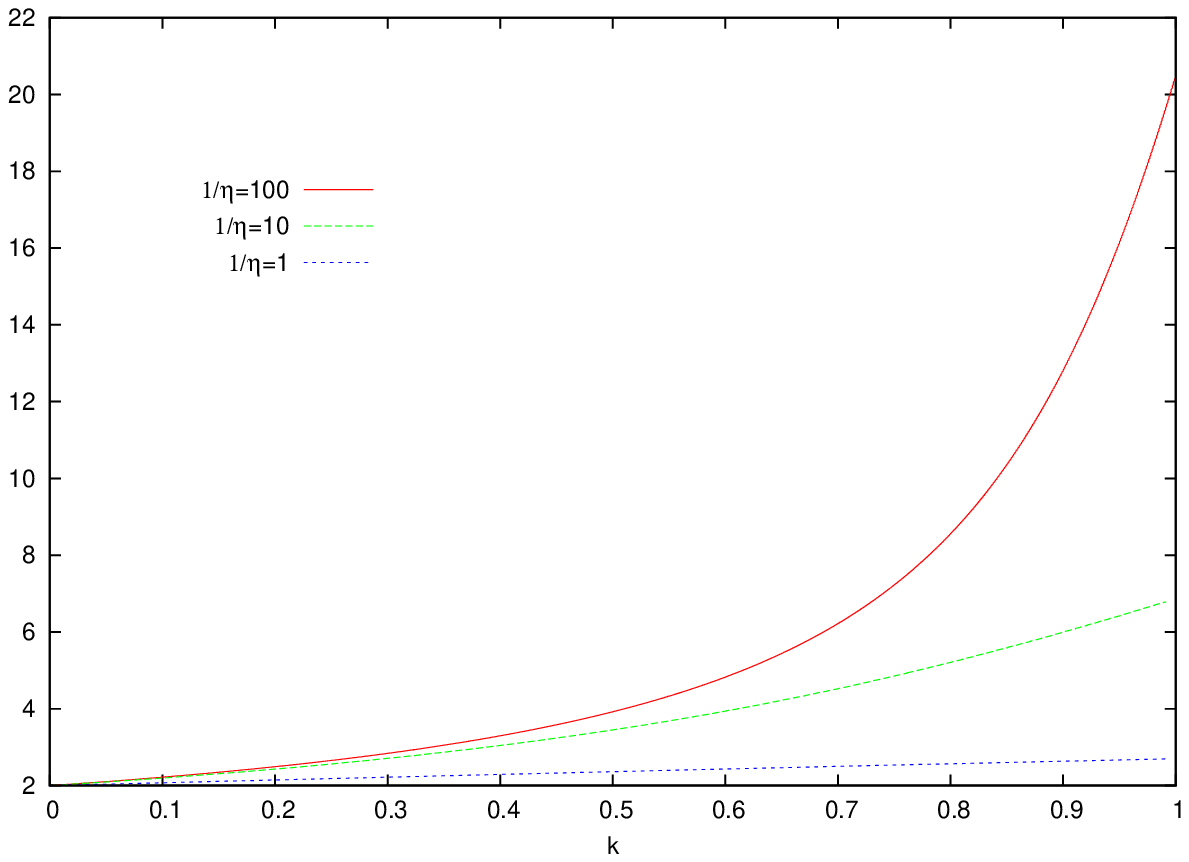}
\end{center}
\caption{\footnotesize
The solution of the steady state
which satisfies
$\alpha S = 2\rho A + kS{\rm e}^{-\eta S/2}$
as function of $1/\eta$ (left) and
$k$ (right).
For simplicity, we set
$\rho=A=1$.
}
 \label{fig:fg5}
\end{figure}
In Fig.  \ref{fig:fg5}, we show
the solution $S$ of the steady state
which satisfies equation (\ref{eq:steady}) 
as function of $1/\eta$ (left) and
$k$ (right).
From this figure, we find that
the $S$ in the steady state increases monotonically in
$1/\eta$ and $k$. 
As we shall discuss in the next section, 
from the view point of maximization of the gross $S$, 
we should increase $k$ and $1/\eta$ to infinity. 
Hence, one cannot choose these parameters
as finite values as $k,1/\eta <\infty$ in the limit of $t \to \infty$.
\subsubsection{Breaking of symmetric phase by noise}
\begin{figure}[ht]
\begin{center}
\includegraphics[width=5.8cm]{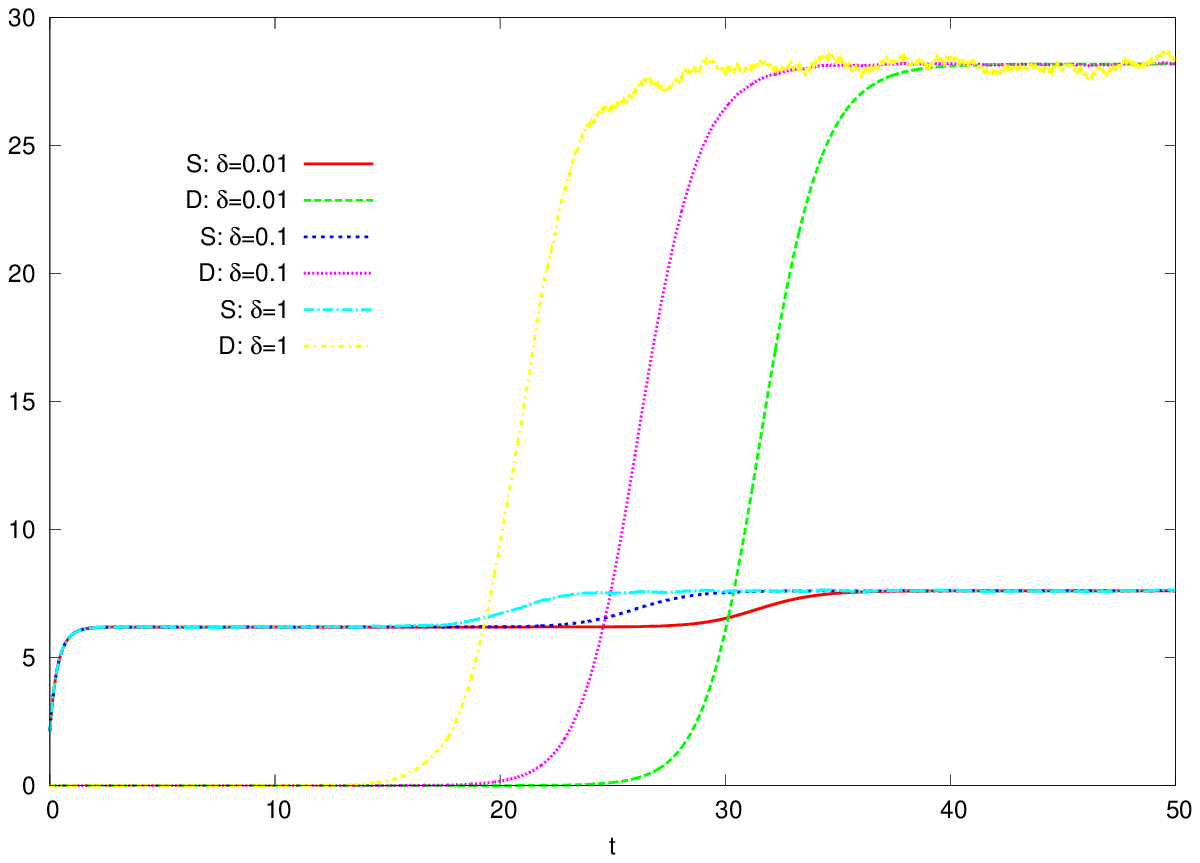}
\includegraphics[width=5.8cm]{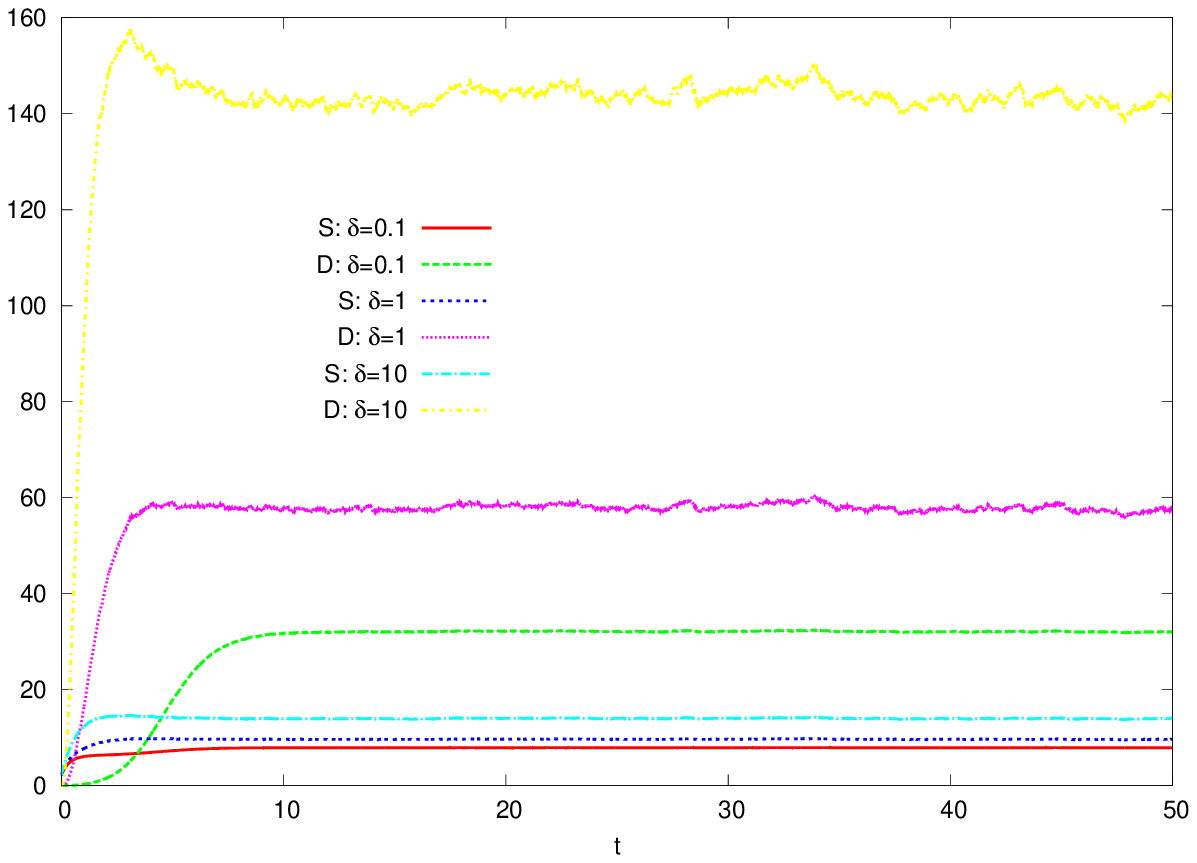} \\
\includegraphics[width=5.8cm]{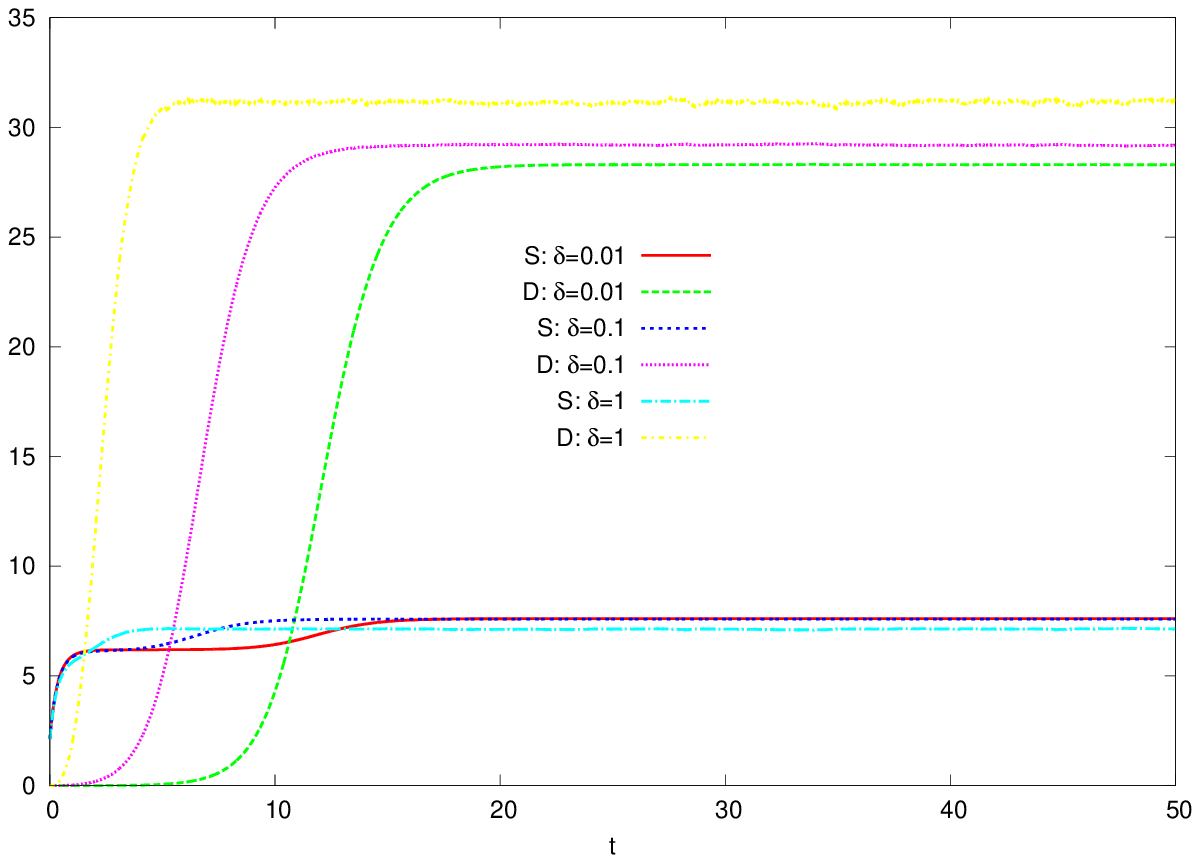}
\includegraphics[width=5.8cm]{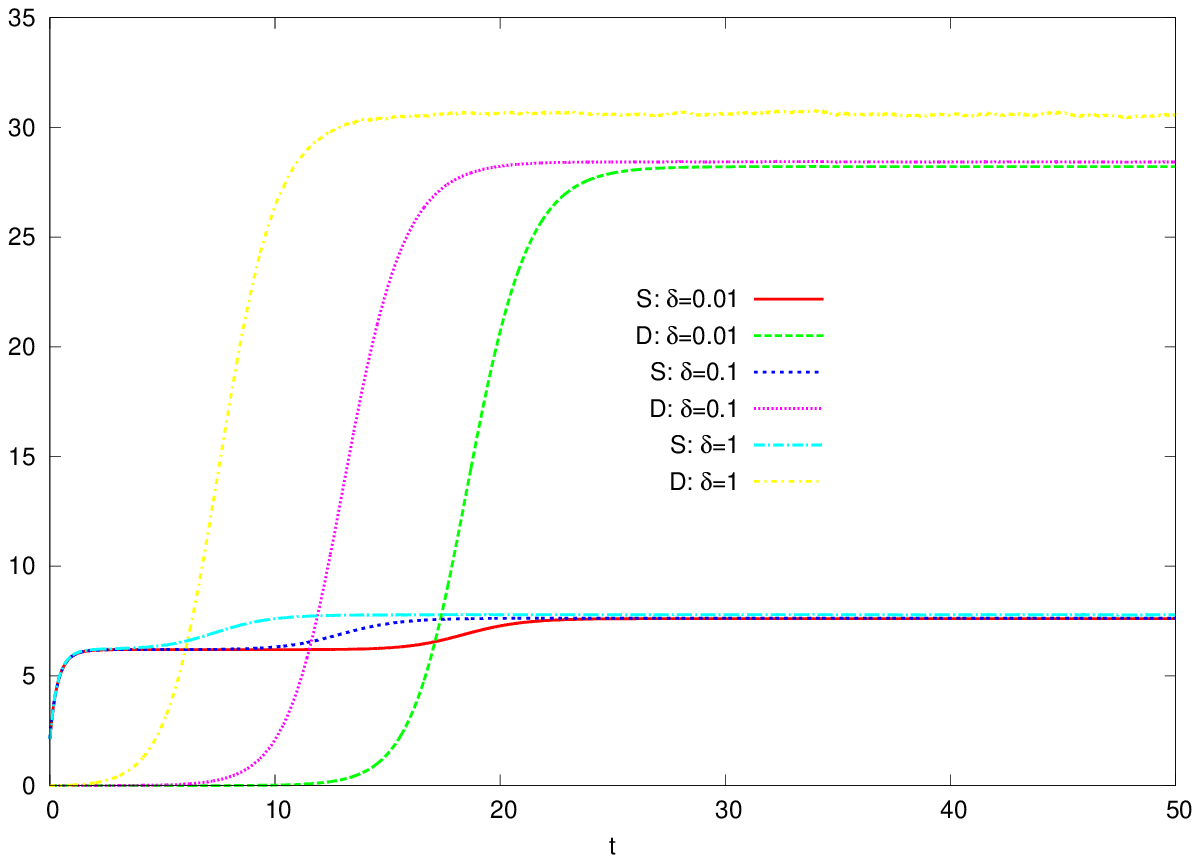}
\end{center}
\caption{\footnotesize
The time-dependence of $S$ and $D$ calculated by Eqs. (\ref{eq:f})-(\ref{eq:g}).
We set the parameters as
$\alpha_2 =\rho_1 =\rho_2 = 1/\eta_2 =1$ and
$k_1 =k_2 =15,
A_1 =A_2= 1$.
Then, we break the symmetry as
$A_{1}=A_{2} +\delta n$ (left) and
$1/\eta_{1}=1/\eta_{2} +\delta |n|$ (right).
The initial condition are selected symmetrically as
$(x_{1}, x_{2})=(1,1)$. }
 \label{fig:fg4}
\end{figure}

As we saw before,
as long as we choose the parameters to satisfy
$A_{1}=A_{2}, \alpha_{1}=\alpha_{2},\cdots$,
we have a symmetric solution $D_{t}=0$.
To break this symmetric phase, here we consider two types of
additive noise, namely:
\begin{enumerate}
\item[(a)] Additive noise on $A_{1}$:
\[
A_{1}=A_{2} +\delta n, \,\,\,n \in [-1,1] \,\,\mbox{(uniform random number)},\,\,A_{2}=1
\]
\item[(b)] Additive noise on $1/\eta_{1}$
\[
1/\eta_{1}=1/\eta_{2} +\delta |n|, \,\,\, n \in[-1,1] \,\,\mbox{(uniform random number)},\,\,\,1/\eta_{2}=1
\]
\end{enumerate}
and change the `amplitude', $\delta$.
The results are shown in
Fig. \ref{fig:fg4}.
In this figure, we set the parameters as
$\alpha_2 =\rho_1 =\rho_2 = 1/\eta_2 =1$ and
$k_1 =k_2 =15,
A_1 =A_2= 1$.
Then, we break the symmetry as
$A_{1}=A_{2} +\delta n$ (left) and
$1/\eta_{1}=1/\eta_{2} +\delta |n|$ (right).
The initial condition are selected symmetrically as
$(x_{1},x_{2})=(1,1)$.

From the left panel, we find that
the symmetric phase specified by
$D_{t}=0$ remains
up to $t_{\rm c}$ even if we add a noise on $A_{1}$.
The $t_{\rm c}$  decreases as the amplitude $\delta$ increases.
On the other hand,
from the right panel,
we find that the symmetric phase is easily broken
when we add a small noise on $1/\eta_{1}$.
In fact, even for $\delta=0.1$,
the critical time $t_{\rm c}$ is close to zero.
Moreover,
we find that
$D_{t}$ rapidly increases when $\delta$ increases and it takes
a maximum at time $t_{\rm p}$.
\section{Optimization of parameters}
In the previous section, we examined
the differential equations
with respect to the gross $S$ and the gap $D$ of
lovers' feelings for some specific choice of parameters $\bm{\theta}$. 

In Rinaldi {\it et. al.} \cite{Rinaldi}, 
they chose those parameters to reproduce 
Scarlett O'Hara and Rhett Butler's feelings according to 
the fascinating plot of the movie GWTW. 
For this purpose, 
the parameters should be time-dependent 
because the fluctuating (up-down) behavior of main characters (Scarlett and Rhett) 
should be induced frequently. 
There is no doubt about this procedure 
of determining the parameters because the GWTH was historically a
remarkable hit movie, 
and the scenario (plot) was the most important factor to make the movie success. 

On the other hand, as explained earlier,
in realistic situations of lovers or married couples, 
the persons do not have to make their feelings fluctuate (up-down) to disrupt their peaceful life and should instead
enhance their psychological uplift so as to keep their affection for the partner strong. 
In this sense, 
we might treat the problem of psychology uplift for the lovers
mathematically, by regarding it as an optimization or an optimal scheduling 
of parameters in the differential equations 
(\ref{eq:f}) and (\ref{eq:g}), so as to 
maximize the gross $S$ and minimize the gap $D$,
as quickly as possible. 

From this viewpoint, we should solve the optimization problem 
for each time step $t$ because the optimal 
parameters are dependent on the time step through the 
gross and the gap. 
Hence, we might choose the variables $\bm{\theta}$, so as to satisfy:
\begin{equation}
\bm{\theta} =
\arg\max_{\bm{\theta}}
\phi(\bm{\theta},\xi: S,D)
\label{eq:hard}
\end{equation}
for each time step, where we defined the following utility function: 
\begin{equation}
\phi(\bm{\theta},\xi: S,D) \equiv
f(\bm{\theta}:S,D)-\xi g(\bm{\theta}:S,D)
\end{equation}
for $\xi \geq 0$. 
The maximization given by equation (\ref{eq:hard}) means 
that we accelerate the speed of increase 
$dS/dt (=f)$ and $-dD/dt (=-g)$ as much as possible 
during the dynamics of $S$ and $D$. 
Hence, when we choose
the variable $S$
for the case of $\xi=0$
to be maximized for lovers, we should optimize the parameters
appearing in the function $f$.
For each time step,
the landscape of $f$ changes due to the dynamics of 
$D$ and $S$, and one should choose
the solution, say, $\bm{\theta}=(\alpha_{1},\alpha_{2},\cdots)$ so as to
maximize the function $f$ at each time step.
As the result, we obtain the trajectory in
the parameter space:
$(\alpha_{1}(0),\alpha_{2}(0),\cdots) \to (\alpha_{1}(1),\alpha_{2}(1),\cdots) \to \cdots$.
\subsection{`Hard' and `soft' optimizations by using a concept of physics}
In the following, for simplicity, we only consider the case of $\xi=0$ 
and we also carry out the maximization of
the speed of increase $dS/dt$ (see equation (\ref{eq:f}) 
and do not take into account the maximization of $-dD/dt$ (see equation (\ref{eq:g})).

To achieve the parameter choice
by means of physics,
we start our argument from the following energy function: 
\begin{equation}
E(\bm{\theta}: S_{t},D_{t}) \equiv
-f(\bm{\theta}: S_{t},D_{t}).
\end{equation}
Obviously, in terms of $f$, we should maximize $f$ as a utility function. 
We should keep in mind that we use the definition of 
$S_{t},D_{t}$ instead of 
$S,D$ to recall us that $f$ is time dependent through those variables. 
We should bear in mind that the function $f$ is defined at each time step $t$. 
In this sense, $f$ is just a function of only parameters $\alpha_{1,2}$, {\it etc.}  
to be selected at each time step. Therefore, the function $f$ is definitely conserved at each time $t$. 

From the viewpoint of `hard optimization', we might utilize the following
gradient descent learning for the parameters $\bm{\theta}$ as
\begin{equation}
\frac{d \bm{\theta}}{dt} =
-\frac{\partial E}{\partial \bm{\theta}} = +
\frac{\partial f}{\partial \bm{\theta}}
\end{equation}
Obviously, the cost function for each time step is dependent on
the state $(S_{t},D_{t})$.
As we mentioned in the previous section (see Fig. \ref{fig:fg4}), $S_{t},D_{t}$ might contain
some noise and through the fluctuation in $S_{t},D_{t}$,
the parameters $\bm{\theta}$ fluctuate
around the peak of the locally concave function $f$. 
To determine the parameters, we temporarily assume here that
the parameters are all `stochastic' variables.
Namely, to adapt ourselves to such realistic cases,
we consider ensemble of the parameters and
we carry out the following maximization of Shannon's entropy under 
the usual two constraints of energy conservation and probability conservation:
\begin{equation}
H =
-\int d \bm{\theta}
P(\bm{\theta}) \log P(\bm{\theta}) + \beta
\left(
E - \int d\bm{\theta} E(\bm{\theta}: S_{t},D_{t}) P(\bm{\theta})
\right) +
\lambda
\left(
1 -
\int d\bm{\theta} P(\bm{\theta})
\right)
\end{equation}
where $\beta, \lambda$ are Lagrange multipliers.
By making use of derivative with respect to $P(\bm{\theta}),\lambda$,
we have
\begin{equation}
P(\bm{\theta}) =
\frac{{\exp}
(\beta f(\bm{\theta}: S_{t},D_{t}))
}
{
\int d \bm{\theta}\,
{\exp}
(\beta f(\bm{\theta}: S_{t},D_{t}))
}
\end{equation}
This is nothing but the Boltzmann-Gibbs distribution
with temperature $T=\beta^{-1}$.

To obtain
the appropriate parameters, we
construct the following
iterations:
\begin{equation}
\bm{\theta}^{(t+1)} =
\int d \bm{\theta}
P(\bm{\theta}) =
\frac{
\int d\bm{\theta}\, \bm{\theta}\,
{\exp}
(\beta f(\bm{\theta}: S_{t},D_{t}))
}
{
\int d \bm{\theta}\,
{\exp}
(\beta f(\bm{\theta}: S_{t},D_{t}))
}
\end{equation}
We should keep in mind that
the strict maximization of $f$ is achieved
by taking the limit of $\beta \to \infty$.
Namely, the solution for `hard optimization' is recovered as
\begin{equation}
\bm{\theta}^{(t+1)}_{\rm hard} =
\lim_{\beta \to \infty}\frac{
\int d\bm{\theta}\, \bm{\theta}\,
{\exp}
(\beta f(\bm{\theta}: S_{t},D_{t}))
}
{
\int d \bm{\theta}\,
{\exp}
(\beta f(\bm{\theta}: S_{t},D_{t}))
}
\label{eq:deterministic}.
\end{equation}
These types of adaptive learning procedure have been well known 
since the reference \cite{Amari} in the literature of neural networks. 

It is important for us to obtain the strict solution, of course. Hoowever, here we consider only the case of $\beta=1$, since we are dealing with the situation in which 
the parameters $\bm{\theta}$ are not deterministic variables; rather, stochastic variables fluctuating around the peaks of $f$.

From the view point of optimization, note that the
function $f$ is not locally `concave' for any choice of $S_{t},D_{t}$.
Hence, the parameters $\bm{\theta}$ which should be selected are trivially going to their `bounds'. 
Nevertheless in the following, we derive the concrete update rule for
each parameter.
We first consider the parameter $\alpha_{1}$.
Here, we assume that $\alpha_{1,2}$ take any value
in $[0,\infty)$.
Hence, we obtain
\begin{equation}
\alpha_{1}^{(t+1)} =
\frac{\int_{0}^{\infty}
d\alpha_{1} \alpha_{1} {\rm e}^{-\alpha_{1} (S_{t}+ \sqrt{D_{t}})/2}}
{\int_{0}^{\infty}
d \alpha_{1} \, {\rm e}^{-\alpha_{1} (S_{t}+ \sqrt{D_{t}})/2}} =
\frac{2}{S_{t}+ \sqrt{D_{t}}},\,\,\,
\alpha_{2}^{(t+1)} =
\frac{2}{S_{t}-\sqrt{D_{t}}}.
\end{equation}
Hence, we find that the parameters $\alpha_{1,2}$ decrease
as inverse of the dynamics $S_{t}$ to zero, when we consider the symmetric case $D_{t}=0$. 
Therefore, as we expected, $\alpha_{1,2}$ go to the bound $\alpha_{1,2}=0$ but the 
optimal scheduling, namely, the speed of convergence to the bound 
$\alpha_{1,2} \sim 1/S_{t}$ is not trivial and would be worthwhile for us to investigate extensively. 

We next consider $\rho_{1,2}$.
For simplicity, we assume that
these two parameters take values in $[0,1]$.
After simple algebra, we have
\begin{equation}
\rho_{1}^{(t+1)}  =
\frac{\int_{0}^{1} d\rho_{1} \rho_{1} {\rm e}^{A_{2}\rho_{1}}}
{\int_{0}^{1}
d\rho_{1}  {\rm e}^{A_{2}\rho_{1}}} =
\frac{A_{2}^{(t)} {\rm e}^{A_{2}^{(t)}} -{\rm e}^{A_{2}^{(t)}} +1}
{A_{2}^{(t)}({\rm e}^{A_{2}^{(t)}}-1)},\,\,\,\,
\rho_{2}^{(t+1)} = \frac{A_{1}^{(t)} {\rm e}^{A_{1}^{(t)}} -{\rm e}^{A_{1}^{(t)}} +1}
{A_{1}^{(t)}({\rm e}^{A_{1}^{(t)}}-1)}.
\end{equation}
Since $\rho$ and $A$ are `conjugates' in the argument of the exponential,
we immediately have
\begin{equation}
A_{1}^{(t+1)} =  \frac{\rho_{2}^{(t)} {\rm e}^{\rho_{2}^{(t)}} -{\rm e}^{\rho_{2}^{(t)}} +1}
{\rho_{2}^{(t)}({\rm e}^{\rho_{2}^{(t)}}-1)},\,\,\,\,
A_{2}^{(t+1)}  =  \frac{\rho_{1}^{(t)} {\rm e}^{\rho_{1}^{(t)}} -{\rm e}^{\rho_{1}^{(t)}} +1}
{\rho_{1}^{(t)}({\rm e}^{\rho_{1}^{(t)}}-1)}.
\end{equation}
For $k_{1,2}$, the structures are exactly similar to those of
$A$ and $\rho$, when we assume that
$k_{1,2}\in [0,1]$.
We easily obtain
\begin{eqnarray}
k_{1}^{(t+1)} & = &
\frac{Q_{1}^{(t)}{\rm e}^{Q_{1}^{(t)}}-{\rm e}^{Q_{1}^{(t)}}+1}
{Q_{1}^{(t)}({\rm e}^{Q_{1}^{(t)}}-1)},\,\,\,
Q_{1}^{(t)} \equiv
\frac{(S_{t}-\sqrt{D_{t}})}{2}{\rm e}^{-\eta_{1}^{(t)}(S_{t}-\sqrt{D_{t}})/2},  \\
k_{2}^{(t+1)} & = &
\frac{Q_{2}^{(t)}{\rm e}^{Q_{2}^{(t)}}-{\rm e}^{Q_{2}^{(t)}}+1}
{Q_{2}^{(t)}({\rm e}^{Q_{2}^{(t)}}-1)},\,\,\,
Q_{2}^{(t)} \equiv
\frac{(S_{t}+\sqrt{D_{t}})}{2}{\rm e}^{-\eta_{2}^{(t)}(S_{t}+\sqrt{D_{t}})/2}.
\end{eqnarray}
Finally, we consider $\eta_{1,2}$.
Here we also assume that $\eta_{1,2} \in [0,\infty)$.
Then, we can write
\begin{eqnarray}
\eta_{1}^{(t+1)} & = &
\frac{\int_{0}^{\infty}
d\eta_{1} \eta_{1}
{\exp}[
\frac{k^{(t)}(S_{t}-\sqrt{D_{t}})}
{2}
{\rm e}^{-\eta_{1} (S_{t}-\sqrt{D_{t}})/2}]}
{\int_{0}^{\infty}
d\eta_{1}
{\exp}[
\frac{k^{(t)}(S_{t}-\sqrt{D_{t}})}
{2}
{\rm e}^{-\eta_{1} (S_{t}-\sqrt{D_{t}})/2}]} \\
\eta_{2}^{(t+1)} & = &
\frac{\int_{0}^{\infty}
d\eta_{2} \eta_{2}
{\exp}[
\frac{k^{(t)}(S_{t}+\sqrt{D_{t}})}
{2}
{\rm e}^{-\eta_{2} (S_{t}+\sqrt{D_{t}})/2}]}
{\int_{0}^{\infty}
d\eta_{2}
{\exp}[
\frac{k^{(t)}(S_{t}-\sqrt{D_{t}})}
{2}
{\rm e}^{-\eta_{1} (S_{t}-\sqrt{D_{t}})/2}]}
\end{eqnarray}
Carrying out the above procedure, one could only `soft' (not `hard') optimize the quantity
$S$.
By substituting the results into
the differential equation with respect to $D$ (see equation (\ref{eq:g})) at the same time,
we may obtain the behavior of the gap.


\section{Discussions and remarks}

In this paper, we first introduced the Rinaldi model and the framework to 
discuss some kind of optimality of a person's behavior, in terms of optimization in the mathematical sense. 
For this, we have just formulated the acceleration rate of 
the gross $dS/dt$, namely, the right hand side of equation (\ref{eq:f}), the function $f$, at each time step. 
However, the function $f$ is not locally concave and the value of optimal parameters go either to 
zero or to infinity, as $t \to \infty$. 
Nevertheless, we can still discuss the scheduling of parameters. For instance, the parameters 
$\alpha_{1,2}$ should decay as $\sim S_{t}^{-1}$ when we attempt to maximize 
the $f$ from the viewpoint of `soft optimization'. 
In near future, we would like to consider and discuss the result of optimization extensively, by considering 
the validity of the model itself. 

Here, we have set $\beta=1$ in the calculations. However, 
we can always regard $\beta$ as a time dependent parameter--  the `inverse-temperature', appearing in the context of `simulated annealing', and  defined by
\begin{equation}
T_{t} = \beta_{t}^{-1}= 
\frac{c_{1}}{(t+c_{2})^{\zeta} (\log (t+c_{3}))^{\epsilon}}
\end{equation}
where the coefficients $c_{1,2,3},\zeta,\epsilon$ determine the speed of convergence. 
As we already mentioned, 
the utility  function $f$ changes through the dynamical variables $S_{t},D_{t}$.
Hence, the utility surface also evolves in time. 
In the above scheduling, we have also assumed that 
the temperate is decreasing within the same time scale 
as dynamical variables $S_{t},D_{t}$ and 
parameters $\bm{\theta}$. However, we can also consider 
the case in which $T$ is scheduled in much shorter time scale than 
$S_{t},D_{t}$ and in the same time scale as $\bm{\theta}$, namely, $T_{\tau}, \bm{\theta}_{\tau}$ with 
$\tau \ll t$. 
Then, the procedure defined by equation (\ref{eq:deterministic}) is regarded as the ``deterministic annealing" \cite{Tishby}. 
In such a general case, 
the optimal scheduling 
for the parameters $\bm{\theta}$ might be changed and 
extensive study along this direction will be reported in our forthcoming paper. 

A few other specific remarks are mentioned below:

\begin{enumerate}
\item In the model considered here, the parameters values are the same as those of Rinaldi \textit{et. al.} \cite{Rinaldi}, but this choice is neither unique nor true for all the ``realistic" situations. A thorough study with other choices of parameters is very much necessary.

\item Identification of the most sensitive parameters responsible for the long time survival of the relationship remains an interesting and open problem. Such identification and then introduction of stochastic fluctuations at the limiting situations could certainly provide more insight towards the modelling approach.

\item The present work is sort of a preliminary attempt of understanding the love dynamics -- theory for the case of sustainability of the love relation between a couple. In reality, the dynamics of love affairs and related modelling approach need more careful and thorough investigations; the effects of several factors have not been considered so far, for example, how the presence of one or more competing person(s) along with the couple, who are in a love relation to each other, can influence the dynamics.
Along the lines of the triangular love studies by Sprott \cite{Sprott1,Sprott2}, it might be very interesting to investigate the role of $S$ and $D$, in order to determine the steady-state relationship between a couple for the case of triangular love. Amongst many other interesting questions, one could also investigate how does a period of separation affect the system dynamics, within this modelling approach.

\end{enumerate}

\begin{acknowledgement}
One of the authors (JI) was financially supported by Grant-in-Aid for Scientific Research (C) of
Japan Society for the Promotion of Science (JSPS) No. 2533027803, 
Grant-in-Aid for Scientific Research (B) No. 26282089, 
and Grant-in-Aid for Scientific Research on Innovative Area No. 2512001313. 
\end{acknowledgement}

%
%

%
%
\end{document}